# DISTOX CALIBRATION TOOLS AND THE NEED FOR CALIBRATION CHECKING

Matija Perne

Društvo za raziskovanje jam Ljubljana, Luize Pesjakove 11, 1000 Ljubljana, Slovenia

Jožef Stefan Institute, Jamova cesta 39, 1000 Ljubljana, Slovenia

matija.perne@ijs.si

For proper cave surveying using DistoX, the device needs to be calibrated with adequate accuracy. Calibrating does not require any tools; but, tools to make calibration easier have been developed. Theoretical consideration shows that the use of certain tools enables one to introduce a type of calibration error that goes undetected by the calibration software. In this study, the existence of such errors is experimentally confirmed and their magnitude is estimated. It is demonstrated to be crucial that the DistoX is calibrated and that the calibration is valid, that is, that the device has not changed since it was last calibrated. No part of the DistoX must have moved or changed its magnetization since calibration, not even the battery. The calibration method used and the quality of the resulting calibration are important too. It is highly recommended that the DistoX be checked immediately before surveying a cave and thus avoid the possibility of using an uncalibrated, not validly calibrated, or poorly calibrated device. To complete the check, a few survey shots are measured multiple times with the device at different roll angles, and the back shot of one of the shots is measured. If the device is properly calibrated, the measurements will agree with each other within the acceptable measurement error. This is not the case for a device that is not properly calibrated.

## INTRODUCTION

DistoX is a device that measures distance, azimuth, and inclination of a survey shot between two stations at a press of a button and can wirelessly transmit the measurements to another device. It is a custom modification of a Disto™ laser distance meter designed for cave surveying (Heeb, 2016; Heeb, 2019a). It is regularly used in cave surveying (Bedford, 2012; Albert, 2017; Kukuljan, 2019; White, 2019) and has been mentioned often in the literature (Gázquez and Calaforra, 2013; Bessone et al., 2016; Ćalić et al., 2016; Mouici et al., 2017; Heggset, 2019) even though its typical use does not result in a publication. It has been used in archaeology (Ortiz et al., 2013; Trimmis, 2018) and mining (Sovero Vargas, 2013). The angular accuracy of the device is reported to be 0.5 degree RMS (Heeb, 2015). And, it compares favorably to the use of a compass and clinometer in cave surveys (Redovniković et al., 2014).

The version DistoX2, studied in this paper, is obtained by replacing the main circuit board of a Leica Disto X310 with a board that contains a STMicroelectronics LIS3LV02DL accelerometer and a PNI Geomagnetic Sensor (STMicroelectronics, 2008; PNI Sensor Corporation, 2016; Heeb, 2019b). These sensors enable the device to measure its own orientation relative to the direction of gravity and the Earth's magnetic field (Heeb, 2009), determining the direction of the laser beam in space.

Manufacturing tolerances and external influences cause measurement errors in determining angles with DistoX. The main errors are eliminated by calibrating the instrument (Heeb, 2008; Heeb, 2009). The calibration procedure consists of measuring multiple unidirectional groups. A unidirectional group is a set of measurements of a fixed but not a priori known direction with various roll angles, turning the device around the beam (Heeb, 2009). Based on these measurements, calibration coefficients are calculated (Heeb, 2015). The coefficients determine a linear correction function that is applied to the sensor values before evaluation (Heeb, 2009). It is recommended to have 14 unidirectional groups of 4 measurements each that are well spread out (Heeb, 2008; Heeb, 2009). After following the recommended procedure, the calibration errors contribute less than 10 % of the total survey error when calibration and survey shots are taken with the same accuracy (Heeb, 2009). Calibration requires no special tools. The user manual remarks that "for best performance, the device should be calibrated in regular intervals" (Heeb, 2015).

The need for regular, precise, time-consuming calibration is an annoyance and mistakes can be made in the process. Several authors using DistoX in their research reassure the reader that the instrument was properly calibrated (Грачев, 2010; Domínguez-Cuesta et al., 2012; Gázquez and Calaforra, 2013; Sovero Vargas, 2013; Pennos et al., 2016; Heggset, 2019). Calibration is one of the reasons for upgrading the DistoX2 to a rechargeable LiPo battery, which eliminates battery replacement as a reason for re-calibrating the device, although the primary reason to use a LiPo battery is to improve compass precision (Heeb, 2014). At a workshop on cave survey organized by Društvo za raziskovanje jam Ljubljana on April 20, 2019, it was noticed that a number of DistoX devices being used in cave surveys reported azimuths that varied by several degrees when the same shot was measured at different roll angles. It became apparent that there was a need for better calibration of the devices.

Several tools that hold the DistoX steady and only allow rotation around the long axis have been developed to make calibration more convenient (Regala, 2016; Kozlov, 2018). It is not clear whether their use results in correct unidirectional groups and in a valid calibration. One of them, the DistoX2 Calibration Cube (Roberson, 2019), was obtained and tested for the study presented in this paper. The version of the tool (current as of September 2019) has two parts. One part is fixed with respect to the ground and consists of a 3D printed cube with holes and three aluminum rod legs. The other part attaches to the DistoX and provides a rod that fits into one of the holes of the cube for every unidirectional group, allowing rotation only around a single axis. Three of the components of the moving part are 3D printed, two are aluminum rods, and one is a brass bolt.

In this study, two DistoX2 devices were calibrated using both the classical procedure from the manual (Heeb, 2008) and the DistoX2 Calibration Cube (Roberson, 2019). The resulting calibrations were tested. It is concluded that it is beneficial to check the DistoX calibration before every use of the instrument.

## THEORETICAL BACKGROUND

A calibrated DistoX uses sensor signals to determine the length, azimuth, and inclination of a survey shot, providing all the measurements required for a cave survey. The distance is measured with a laser distance meter. The azimuth and inclination are deduced from the direction of gravity and the direction of the Earth's magnetic field relative to the laser beam. The direction of gravity and of the magnetic field are calculated from the signals of built-in accelerometers and geomagnetic sensors. These signals are affected by systematic measurement errors, such as those resulting from the following (Heeb, 2009):

- the offset and gain errors of the sensors,
- incorrect mounting angles of the magnetic sensors and accelerometers in relation to one another and the laser beam,
- influence of the metallic parts of the instrument, such as the battery, on the magnetic field.

These errors can be eliminated with a linear function, and the coefficients of the necessary transformation can be calculated from a series of calibration measurements (Heeb, 2009).

The correction to gravity is applied with the formula

$$\vec{gr} = \boldsymbol{G}\,\vec{gs} + \vec{gd}$$

in which $\vec{gr}$ is the resulting gravity and is a vector with three components, $\boldsymbol{G}$ is a 3 by 3 transformation matrix, $\vec{gs}$ is the vector of the sensor values, and $\vec{gd}$ is the offset. The transformation is determined by 12 coefficients, that is, 9 elements of $\boldsymbol{G}$ and 3 elements of $\vec{gd}$, that have to be calculated from calibration shots. By convention, the $x$ direction of the coordinate system of $\vec{gr}$ is pointing along the laser beam, $y$ is to the right and $z$ down (Fig. 1).

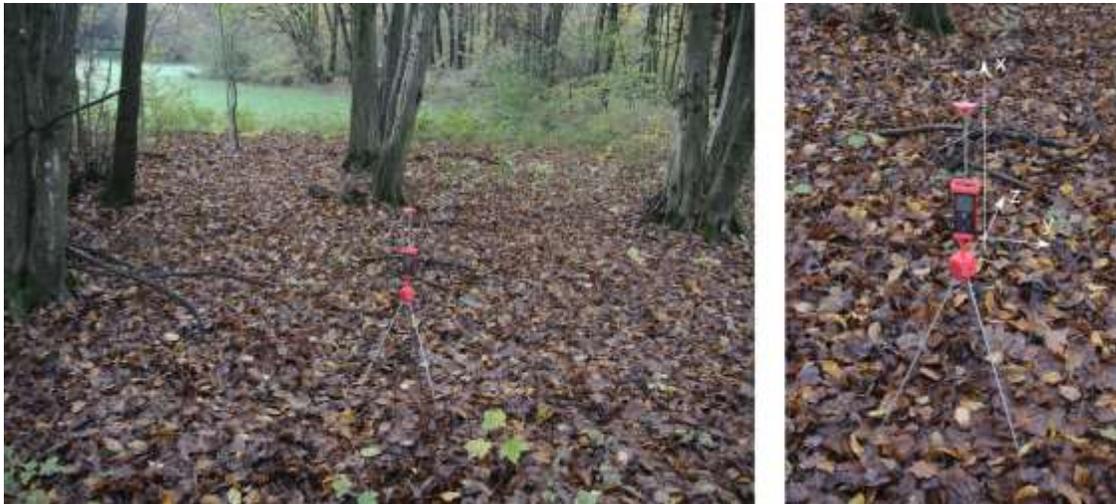

Figure 1. DistoX on the calibration tool in the field work location. The coordinate system is defined as in Heeb (2009).

A rigorous analysis of computing the calibration coefficients from various types of calibration shots is given in Heeb (2009). Heeb analyzed the three following possibilities:

- using shots of known directions,
- using random free measurements of unknown directions, and
- using unidirectional groups of several shots in the same direction at different roll angles.

Heeb determined that the known directions method is impractical because stations in known relative positions are typically not available. And, free measurements are not sufficient for determining the angles between the sensors and the laser beam. The method used in practice is the unidirectional group method. Unidirectional group measurements are sufficient for determining all of the calibration coefficients except for one parameter that is related to the roll angle and is ambiguous. It does not influence the use of the device though and the choice $G_{yz} = G_{zy}$ is made to set it (Heeb, 2009).

The equations for the magnetic field are of the same form as the equations for gravity and are used the same way.

According to Heeb (2009), 4 evenly spread out unidirectional groups of 4 shots at different roll angles, combined with free measurements, are enough to determine the calibration coefficients with a good precision. More groups increase the precision; so, 14 unidirectional groups of four shots with evenly spread out roll angles around the beam are recommended. The coefficients are determined with an optimization method from all the shots, averaging out the random errors.

Corvi (2017) reports that 24 or 27 coefficients result from calibration. The 12 for gravity and the 12 for magnetism add up to 24, while the extra three are for the nonlinear correction, if used. It was determined that nonlinearity of the accelerometers may cause a significant systematic error after linear calibration, so support for a simple second order correction function was added to the firmware (Heeb, 2014).

### When to re-calibrate the DistoX?

When any source of error that is corrected by the calibration changes in size, the calibration coefficients cease to be valid and the instrument must be re-calibrated. The parts of the instrument must not move relative to one another in order for the calibration coefficients to be constant. If the device is jostled enough that critical parts move without getting loose, re-calibration will help.

The influence of the device's metallic parts on the magnetic field can change if the parts either physically move or change their magnetization. A re-calibration after a battery change is unavoidable (Heeb, 2009); any movement of the batteries may change the coefficients. A serious magnetization of the ferromagnetic material in the instrument by an invisible magnetic field can harm the instrument's precision (Heeb, 2016) and a magnetization change orders of magnitude smaller would suffice to noticeably change the calibration coefficients.

Due to component drift and aging, the offset and gain errors of the sensors change with time and calibration has to be repeated occasionally (Heeb, 2009).

Travelling for a long distance does not necessitate re-calibration in itself though. The relationship between the external fields and the sensor values is independent from the location. A calibrated device works equally well everywhere in the world (Heeb, 2009), even though gravity, the strength of the geomagnetic field, and the magnetic inclination are different in different places. Magnetic declination, however, is a separate and unrelated issue. Devices like DistoX measure the magnetic azimuth. When a different azimuth is needed, conversion is necessary.

*Quality measures and their limitations*

When the calibration coefficients are calculated from the unidirectional groups, it is checked how accurately the roll axes $x$ of the shots within each group point in the same direction. If the spread is big, the calibration coefficients are likely not accurate. The calibration software computes several quality measures that quantify the average error. An automated warning is provided if certain ones are above a set threshold (Corvi, 2020).

The laser beam is assumed to be pointing along the $x$ axis (Heeb, 2009). If, due to random errors, the laser is not pointing exactly in the same direction in all of the shots within each unidirectional group, the algorithm will be less precise in matching the $x$ axis and the beam. At the same time, the determined $x$ axis will be found to not have pointed in a constant direction within each unidirectional group, so the quality measure will be big.

A calibration tool provides an $x$ axis that is independent from the laser beam. Correct calibration with a tool depends on the match between the $x$ axis provided by the tool and the laser beam. The mismatch between the two does not contribute to the quality measure – as long as the $x$ axis is pointing in the same direction in all the shots of each group, the quality measure will be small. In the case of a stable $x$ axis and a poor match between it and the laser beam, a poor calibration with a good quality measure and no warning will result. The mismatch angle is likely to be dependent on the device and the calibration tool and may vary between calibrations.

Some of the calibration tools, for example the Calibration Cube, require attaching parts to the DistoX for the duration of the calibration. The instructions correctly specify that all the parts should be non-magnetic – plastic, aluminum, brass (Roberson, 2019) – but a slight lapse of attention could result in the use of e.g. a steel washer. The calibration would faithfully include its influence on the magnetic environment in the instrument, so the quality measure would be small. However, the calibration coefficients would only be valid until the magnetic part is removed.

**EXPERIMENTAL**

To test the difference between the calibrations using the method described in the manual and using the Calibration Cube, the following procedure was followed:
1. obtain a DistoX
2. take it to a wooded area with an even magnetic environment
3. test the DistoX on a triangular test course, surveying in both directions, measuring each shot four times with evenly spread roll angles
4. calibrate the DistoX with the Calibration Cube
5. repeat the test under number 3
6. calibrate the DistoX following the DistoX Calibration Manual (Heeb, 2008)
7. repeat the test under number 3
8. repeat some of the points 4–7 depending on the collected data
9. go back to 1 if necessary.

First, a Calibration Cube had to be acquired. The holes of the 3D model were changed into metric sizes and the model was printed.

As soon as a DistoX was obtained, a mistake in the assumptions became apparent. The assumption was that the brass screw of the Calibration Cube that holds the DistoX does so by pinching the device. Therefore, the hole in the 3D printed part was increased in diameter to 8 mm and another 4 mm hole was added next to it, reflecting the sizes of brass bolts at hand. However, the bolt is supposed to engage an internal 1/4-inch thread in the DistoX. In the first tests, the DistoX was pinched, while for the later ones, a custom brass bolt was made – 1/4-inch brass bolts are indeed not easily available in Slovenia where the work was performed – and the hole in the tool was made smaller with several layers of nail polish (Fig. 2). For this reason, we have not tested the calibration tool exactly as it was designed, but it functioned as intended.

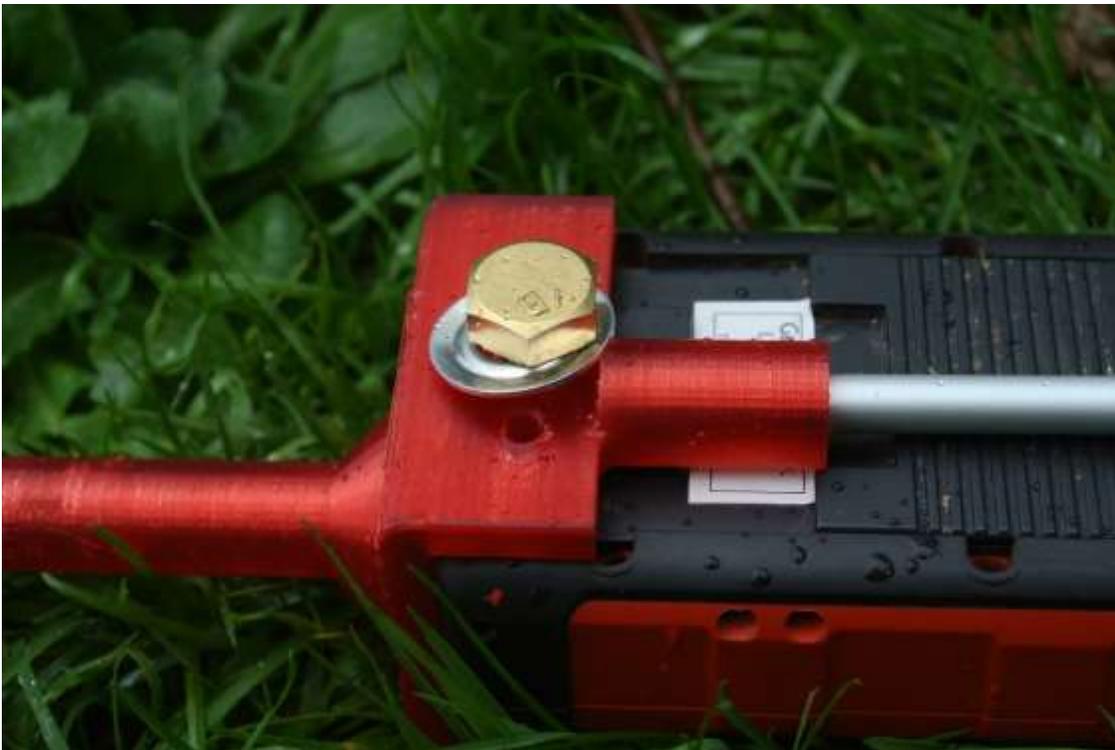

Figure 2. Attaching the DistoX to the calibration tool with an almost correct bolt. The bolt is slightly too long and we avoided sawing it by adding an aluminum washer. The hole on the tool, which was originally too big, was shrunk using a few layers of nail polish.

The field work was performed with two DistoX devices in a forest close to location 33 T 447502 E 5113134 N (WGS 84 46.169761 14.319938). Three test course stations were marked on tree trunks 5 to 8 m apart and at inclinations under 30° from one another. The Cube calibrations were done in the immediate vicinity (Fig. 1) and the Cube was always moved between successive calibrations. The classical calibrations were performed in the immediate surroundings as well, following the DistoX Calibration Manual of Heeb (2008). The unidirectional groups of shots were taken between tree trunks, from the tree trunks to the ground, and from the ground to the trees. As several of the calibration stations were temporary features or out of reach, they were not documented and not systematically reused from one calibration to another. No part of the test course was used in any calibration. Both tested devices use a built-in LiPo battery.

The DistoX work was carried out over four separate days and at different times of day. Several technical issues bothered us on the first day, including too much ambient light to see the laser dot easily, not enough light to see the target easily, and DistoX pinched to the Calibration Cube using a 4 mm bolt. The majority of the field work was performed on the following three days, after resolving the technical issues, with optimal equipment and workforce and mostly optimal weather. All of the work was done with care, none of it was performed when the circumstances felt unsuitable for accurate work. Different calibration methods and their tests were performed in a random fashion with the number of classical and Cube calibrations and the numbers of tests of both balanced on each day of field work. Any effect of the date on the measurement accuracy would thus not correlate with the calibration method. The log with the details is provided with the data set (Perne 2020a).

The azimuth and inclination angles of the test course were checked with a Suunto handheld compass and clinometer so that the DistoX results can be compared with values obtained with an independent method.

The field experience has shown that the central cube of the tool is very stable when the legs are stuck in the soil (Fig. 1). In the Calibration Cube created for this study, the DistoX-containing moving part is less stable than the central cube and wobbles significantly. Care was taken to minimize the wobble by supporting the DistoX with hands but it was not clear how much of the wobble remained.

A smartphone with Android and TopoDroid version 4.1.4G was used for computing the calibration coefficients from the calibration shots and for recording the test data.

**DATA ANALYSIS AND RESULTS**

Raw measurement data with detailed metadata and the code used for calculating the values presented in the tables and the text is freely available in the online data set of Perne (2020a). Data processing is done in R (R Core Team, 2020) using the package dplyr (Wickham, 2020).

### Data exploration

Table 1 contains several statistics of DistoX tests. The most important numbers are reproduced in the graph in Figure 3. Table 2 presents the data on the instrument calibrations themselves.

**Table 1. Results of the tests of the calibrations. All the σ's and the Error stddev are in angular degrees, all the Δ's are in meters. Each row represents one test, the more or less strong yellow ones are tests of the calibrations with the tool, the red ones refer to the classical calibrations from one tree to the other. When neighboring lines are of the same shade, they test the same calibration. We measured six test shots (a triangle in both directions), each one four times with the DistoX in different orientations. In the columns <σₐ> and <σ_φ> are the average standard deviations in the azimuth and the inclination, averaged over the six shots. The columns max(σₐ), max(σ_φ), min(σₐ) and min(σ_φ) contain the biggest and the smallest standard deviations of both angles picked from the data on the six shots. The columns rmsΔtot, rmsΔvert and rmsΔhor are the root-mean-square mismatches of the closure of the triangle: total, vertical, and horizontal, respectively. As each side is measured eight times (four times in each direction), we get 8³ = 512 possible surveys of the triangle and average over them all to obtain these numbers. The remaining columns are the total, vertical, and horizontal closure mismatches of the averaged triangle survey. The table is generated by the script "DistoX.R" (Perne, 2020a).**

| Test | <σₐ> | <σ_φ> | max(σₐ) | max(σ_φ) | min(σₐ) | min(σ_φ) | rmsΔtot | rmsΔvert | rmsΔhor | Δtot | Δvert | Δhor | Calibration Error stddev |
|------|------|-------|---------|---------|---------|---------|---------|----------|---------|------|-------|------|--------------------------|
| n1original | 1.717 | 0.255 | 3.021 | 0.386 | 0.685 | 0.171 | 0.289 | 0.047 | 0.285 | 0.029 | 0.010 | 0.028 | |
| n2original | 3.402 | 0.629 | 5.690 | 0.730 | 1.061 | 0.556 | 0.697 | 0.107 | 0.688 | 0.037 | -0.006 | 0.037 | |
| n1k1t1p | 0.390 | 0.288 | 0.532 | 0.419 | 0.216 | 0.173 | 0.128 | 0.087 | 0.094 | 0.047 | -0.027 | 0.038 | 0.24 |
| n1k1t2p | 0.364 | 0.321 | 0.457 | 0.443 | 0.200 | 0.250 | 0.119 | 0.071 | 0.095 | 0.031 | -0.010 | 0.030 | 0.24 |
| n1k1t3p | 0.369 | 0.274 | 0.658 | 0.311 | 0.222 | 0.222 | 0.109 | 0.063 | 0.089 | 0.021 | -0.001 | 0.021 | 0.24 |
| n1k2t1p | 0.458 | 0.403 | 0.915 | 0.465 | 0.208 | 0.356 | 0.138 | 0.083 | 0.110 | 0.028 | -0.010 | 0.026 | 0.18 |
| n1k2t2 | 0.477 | 0.390 | 0.835 | 0.443 | 0.250 | 0.346 | 0.123 | 0.074 | 0.098 | 0.014 | 0.001 | 0.014 | 0.18 |
| n2k1t1 | 0.853 | 0.786 | 0.954 | 0.826 | 0.751 | 0.735 | 0.200 | 0.133 | 0.150 | 0.022 | 0.014 | 0.016 | 0.17 |
| n1g1t1 | 0.282 | 0.187 | 0.574 | 0.299 | 0.058 | 0.096 | 0.117 | 0.084 | 0.082 | 0.027 | 0.013 | 0.023 | 0.18 |
| n1g1t2p | 0.236 | 0.127 | 0.379 | 0.171 | 0.096 | 0.082 | 0.105 | 0.065 | 0.083 | 0.021 | 0.019 | 0.009 | 0.18 |
| n1g1t3 | 0.264 | 0.110 | 0.392 | 0.183 | 0.129 | 0.058 | 0.092 | 0.049 | 0.077 | 0.015 | -0.001 | 0.015 | 0.18 |
| n1g2t1 | 0.316 | 0.124 | 0.457 | 0.150 | 0.189 | 0.082 | 0.084 | 0.050 | 0.067 | 0.019 | 0.009 | 0.016 | 0.29 |
| n2g1t1 | 0.163 | 0.057 | 0.299 | 0.096 | 0.082 | 0.000 | 0.041 | 0.018 | 0.037 | 0.020 | -0.002 | 0.020 | 0.05 |

**Table 2. TopoDroid's criticism of the calibrations. According to the criteria of the app, all are reasonably good. It seems that the wobble of the DistoX on the Calibration Cube is small enough that it does not harm the result.**

| Name | Average error [°] | BH delta | Error stddev [°] | Max. error [°] | Iterations |
|------|-------------------|----------|------------------|----------------|------------|
| n1g1 | 0.28 | 0.566 | 0.18 | 1.01 | 39 |
| n1g2 | 0.28 | 0.806 | 0.29 | 2.45 | 36 |
| n1k1 | 0.36 | 0.688 | 0.24 | 1.18 | 34 |
| n1k2 | 0.29 | 0.555 | 0.18 | 1.01 | 34 |
| n2g1 | 0.1 | 0.214 | 0.05 | 0.23 | 31 |
| n2k1 | 0.25 | 0.438 | 0.17 | 0.78 | 28 |

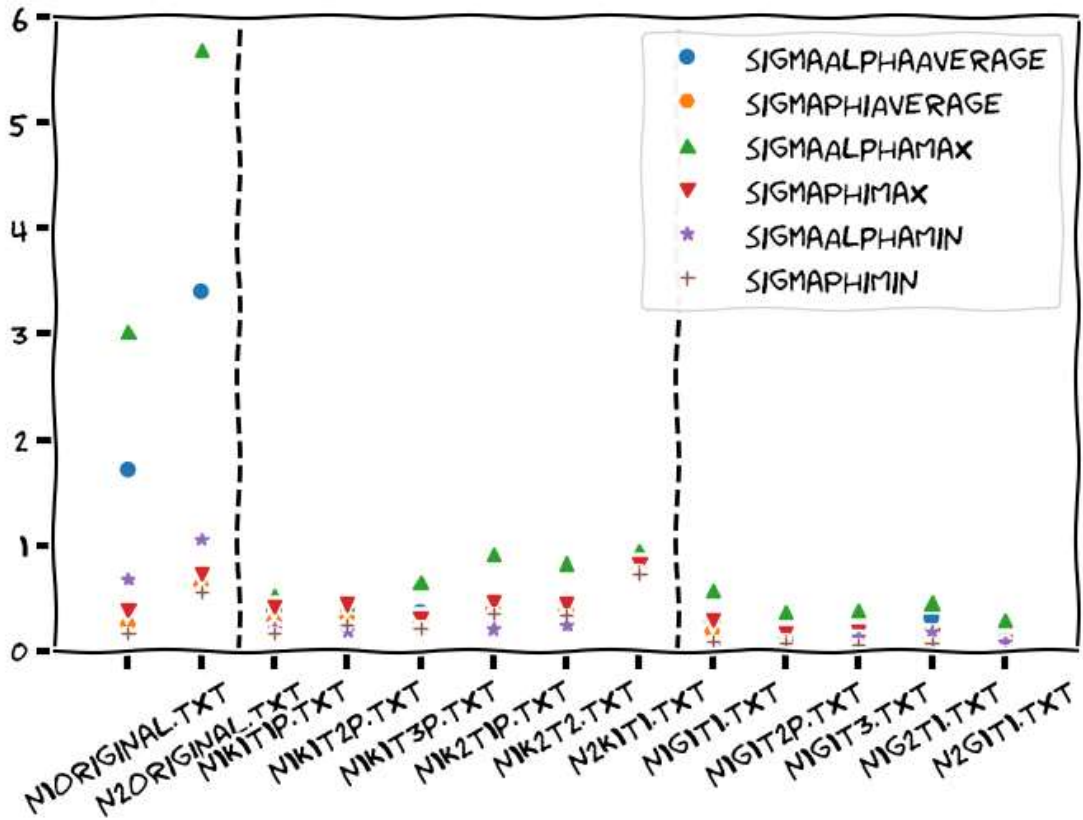

Figure 3. Graphical representation of the σ columns of Table 1.

The names of the calibration files (Table 2) start with the label of the device, which is either "n1" or "n2". The label of the calibration follows, consisting of the letter "k" for a Calibration Cube calibration or the letter "g" for the classical calibration, followed by the consecutive number of the given type of calibration on the particular device. The names of the test files (Table 1) start with the name of the calibration file, which is followed by the letter "t" and the consecutive number of the test of the given calibration. For the files that required a manual correction of a mistake in the raw data, the letter "p" follows, and all the manual corrections are documented in the data set. The file name structure is different for the initial testing of the preexisting calibration of each instrument, and consists of the label of the device followed by the word "original".

### *Statistical analysis*

A rigorous statistical analysis is a reliable and accurate way of quantifying the studied effects and is complementary with data exploration.

The measured angle value in a single test shot is written as
$$\hat{\alpha} = \alpha + \nu,$$
where $\hat{\alpha}$ is the measured value of the angle (azimuth or inclination), $\alpha$ is the true value, and $\nu$ is the measurement noise. Noise is a random variable that is sampled from an unknown distribution. The possible hypotheses for the measurement noise of differently calibrated DistoXs are the following:

1. the noise distributions have the same variance for both calibration methods;
2. the devices calibrated either way are equally accurate and any difference in the variance of the noise distribution arises during testing from causes unrelated to calibration;
3. the calibrations are different and influence the variance of the noise distribution observed by the tests but the difference in the calibration method is not the cause, both calibration methods are equally good;
4. the Cube calibration method is not as good as the classical one, increasing the variance observed in the tests.

It will be shown that hypotheses 1 to 3 can be rejected.

To measure the noise, the true value has to be known, but it is not. It is thus approximated with a mean value. Five different mean values are considered:

1. arithmetic mean of all the measurements of a shot over all tests;
2. weighted arithmetic mean of all the measurements of a shot, where the weight is the inverse of the variance of all the shot angles in a given test;
3. arithmetic mean of the four measurements of a shot in the particular test;
4. average of the arithmetic mean and the inverted arithmetic mean of the back shot;
5. average of the weighted arithmetic mean and the inverted weighted arithmetic mean of the back shot.

Each approximation ascribes a different meaning to the true value and it is not evident which one is the most meaningful. The calculations are repeated with all of them and the full results are reported for the approximate true value no. 2. because it is the most intuitive one.

Hypothesis 1 is tested statistically with the modified robust Brown-Forsythe Levene-type test based on the absolute deviations from the median as implemented in levene.test function of Gastwirth et al. (2020). The noise variance in the 5 tests of classical calibrations is compared to the one in the 6 tests of Cube calibrations, and the $p$-value for the azimuth is found to be $3.6 \cdot 10^{-6}$. According to the same Brown-Forsythe test, the noise standard deviation in the tests of the Cube calibrations is most likely 1.6 times larger than the noise standard deviation in the tests of the classical calibrations. The ratio of the noise standard deviations is at least 1.35 at 95 % confidence level. For inclination, the equivalent $p$-value is $1.3 \cdot 10^{-11}$, the most likely ratio is 2.2, and the ratio is 1.85 or above at 95% confidence level. The result is robust with respect to the choice of the mean value used to approximate the true value; at 95% confidence, the noise standard deviation ratio is 1.22 or above in azimuth and 1.84 or above in inclination with every mean value used. Hypothesis 1 that the noise distributions have equal variances is thus firmly rejected and it is shown that the difference in variances is not small.

When the calibration tests are ranked in DistoX performance (Table 1), all 5 tests of classical calibrations are better than any of the 6 tests of the calibrations with the Cube. The claim holds true for every angle deviation measure apart from the maximum[1]. Only one combination in $\binom{11}{5} = 462$ behaves this way, making the occurrence unlikely if the DistoX and its calibration had no influence on the test results. Hypothesis 2 is thus rejected.

---

[1] The maximum is the most sensitve to outliers and thus the least likely to give the true picture.

The 3 classical calibrations perform better in testing than any of the 3 with the Cube. The probability of this happening if the calibration method had no influence on calibration quality is one in $\binom{6}{3} = 20$, which is too high to confidently reject hypothesis 3. However, calibration is designed to contribute less than 10 % of the survey error (Heeb, 2009) so its influence should be below the detection limit of the tests. The observed difference is too large. Hypothesis 3 that the calibration methods are both good but the Cube resulted in worse calibrations by chance is therefore rejected as an explanation for the observed result.

It follows that the remaining hypothesis 4 that the Cube leads to worse calibration than the classical method should be accepted.

Validation against other measurement devices

The survey of the test course with handheld compass and clinometer and the code comparing it with the DistoX measurements is provided in the data set (Perne, 2020a). No significant difference between DistoX and Suunto angles is observed. The differences between the mean values of each method are within 1 angular degree, which is within the measurement error of Suunto instruments. The systematic error of DistoX compared to Suunto is not detectable.

**DISCUSSION OF THE RESULTS**

The statistical tests show that the observed Cube calibrations negatively influence the performance of DistoX compared to classical calibrations but are not recognized as poor by the calibration quality measures. The finding can be explained with the theoretical prediction of the consequences of providing an $x$ axis independent from the laser beam.

Inspecting the test results (Table 1 and Fig. 3), the calibration that performs worst is n2k1, a Calibration Cube calibration. Its standard deviations of the angles are very similar for every test shot and in both azimuth and inclination, the difference between $\max(\sigma_\alpha)$ and $\min(\sigma_\varphi)$ is small. This is the expected error pattern if the DistoX follows a conical surface in each shot when the roll angle changes, as if the calibration $x$ axis was not parallel to the beam. It is thus in good agreement with the predicted side effect of the use of calibration tools.

As expected, the errors of Calibration Cube calibrations do not influence the quality measures (Table 2), some Cube calibrations are graded better than some classical calibrations but perform worse in tests. This is because the error cannot be detected in the calibration shots if it results from the $x$ axis being at an angle to the beam.

The data on the mismatch of the triangles is provided to illustrate the influence of calibration on cave surveys. It is interesting to see how the errors average out if the DistoX is being rotated.

The DistoX owners were not asked about the calibration state of their devices, reflecting the fact that the devices would likely be assumed to be calibrated if borrowed for cave surveying purposes. The owner of n1 mentioned that he compared it with a compass and was not impressed, while n2 was reportedly being used as if it was calibrated well. According to the test results (Table 1), both devices when borrowed seem to have achieved UIS survey grade 3 (Häuselmann, 2011) accuracy but n2 seems not to have met BCRA survey grade 3 (British Cave Research Association, 2002) standards.

The calibration n1g2 was performed very carefully with the goal of achieving similar results as in the excellent n2g1, but it did not work out regarding the calibration quality measures. It does not seem to be a result of human error. More likely it is a consequence of a subtle difference between the DistoXs, perhaps just a fussy measure button. The calibration n1g2 nevertheless achieved acceptable quality measures and performed flawlessly in the test.

### Practical implications of the results

One should regularly check the calibration of a DistoX used for cave surveying. Inferring that it is calibrated well based on its history is not reliable because one may not have the complete information. Calibration should be checked at least at the beginning of every survey so that a possible problem is detected before causing damage. It may be beneficial to check the calibration again at the end of the survey to better constrain the quality of the device throughout the survey work. One may even vary the roll angle whenever taking a multiple measurement of a survey shot in order to detect any issues as quickly as possible, although it is not clear whether the benefit outweighs the additional cost.

One should not use a DistoX calibration tool without checking the calibration before use. A tool increases the likelihood of substantial calibration errors that are not detected by the calibration quality measures, so a calibration check is the only way of detecting them.

When the same shot is measured multiple times at evenly spread out roll angles, i.e., with the display of the DistoX pointing in different directions, the measurement result variation should be within the desired survey accuracy. The calibration can be checked by measuring several shots with varying roll angles and measuring a shot in both directions, checking the agreement of the values. The test with the back shot is necessary because changing the roll does not detect offsets in the $x$ direction.[2] A rigorous derivation of the necessary and sufficient conditions for the correctness of the calibration is given by Corvi (2018). This calibration check method has been proposed before. The Calibration Manual (Heeb, 2008) recommends performing the procedure after calibration as a quick check. TopoDroid User Manual (Corvi, 2020) mentions that calibration-check shots should be taken at different roll angles and describes the functionality of TopoDroid related to check shots. However, the need for regular calibration checks seems to not have been emphasized enough.

---

[2] The author thanks Beat Heeb for pointing it out while taking full responsibility for the claim.

The recommendation of checking the calibration with check survey shots is not contradicted by the claim that calibration only contributes up to 10 % to the total survey error (Heeb, 2009). This type of calibration check cannot see the difference between an excellent and a mediocre calibration, but it can detect one that is so poor that it would importantly influence the survey accuracy.

If a survey is performed with a poorly calibrated DistoX, one may want to calibrate the device and correct the calibration errors after the survey. The survey azimuth and inclination values are not enough to make it possible because the conversion of the 6 raw sensor values into the 2 angles is irreversible. Additional information on the sensor readings or DistoX orientation is necessary for a successful correction.

**CONCLUSIONS**

It is crucial for a DistoX in cave survey use to have a fresh, valid calibration. The quality of the calibration, which depends on the method used, is somewhat important as well. Calibration tools offer opportunities for making the rotation axis not parallel to the laser beam or for having a magnetic part attached to the instrument during calibration, increasing the calibration errors. These errors are not detected in the computation of the calibration coefficients. Such mistakes do not occur if the calibration manual is followed literally and no tool is used. Nevertheless, several mechanisms that can render calibration coefficients invalid unbeknownst to the user do exist regardless of the calibration method. Regular checking of the calibration, at least at the beginning of every survey, is thus highly recommended. If calibration is checked regularly, use of calibration tools may not be problematic as any sizeable calibration error would be detected and corrected before causing damage.

**NOTATIONS LIST**

| | |
|---|---|
| $\langle \cdot \rangle$ | Average |
| $\boldsymbol{G}$ | Matrix of calibration coefficients [-] |
| $\overrightarrow{gd}$ | Offset of accelerometers [m/s$^2$] |
| $\overrightarrow{gr}$ | Resulting gravity [m/s$^2$] |
| $\overrightarrow{gs}$ | Sensor signal of gravity [m/s$^2$] |
| $\max(\cdot)$ | Maximum |
| $\min(\cdot)$ | Minimum |
| rms$\Delta_{hor}$ | Horizontal root-mean-square mismatch of the triangle closure [m] |
| rms$\Delta_{tot}$, | Total root-mean-square mismatch of the triangle closure [m] |
| rms$\Delta_{vert}$ | Vertical root-mean-square mismatch of the triangle closure [m] |
| $x$ | Coordinate along the laser beam |
| $y$ | Coordinate to the right with respect to the DistoX |

| $z$ | Coordinate down with respect to the DistoX |
| $\Delta_{hor}$ | Horizontal root-mean-square mismatch of the closure of the average triangle [m] |
| $\Delta_{tot}$, | Total root-mean-square mismatch of the closure of the average triangle [m] |
| $\Delta_{vert}$ | Vertical root-mean-square mismatch of the closure of the average triangle [m] |
| α | true value of an angle (azimuth or inclination) [°] |
| $\hat{\alpha}$ | measured value of an angle [°] |
| ν | measurement noise [°] |
| $\sigma_\alpha$ | Standard deviation in azimuth for a shot [°] |
| $\sigma_\varphi$ | Standard deviation in inclination for a shot [°] |

**ACKNOWLEDGMENTS**

Marjan Baričič, Matic Di Batista, Gregor Pintar, Peter Prevec, Janez Strojan, Stephanie Sullivan and Rafko Urankar-Cile assisted in completing the work. Društvo za raziskovanje jam Ljubljana and its journal Glas podzemlja have allowed the article Perne (2020b) to be translated, modified and extended. I sincerely thank Paul Burger, Marco Corvi, and Beat Heeb for substantial comments on earlier drafts of the manuscript. Editing and revisioning of the article was financially supported by the Slovenian Research Agency through research core funding No. P2-0001.